\documentclass[12pt]{article}
\usepackage{graphicx}
\usepackage{epsfig}
\def\beq{\begin{equation}}
\def\eeq{\end{equation}}
\def\bea{\begin{eqnarray}}
\def\eea{\end{eqnarray}}
\def\nn{\nonumber}
\def\sss{\scriptscriptstyle}

\def\roughly#1{\mathrel{\raise.3ex\hbox
{$#1$\kern-.75em\lower1ex\hbox{$\sim$}}}}

\def\gsim{\roughly>}
\def\sla#1{\raise.15ex\hbox{$/$}\kern-.57em #1}
\def\bra#1{\left\langle #1\right|}
\def\ket#1{\left| #1\right\rangle}
\def\ks{K_{\sss S}}
\def \Bbar{\bar B}
\def \fbar{\bar{f}}
\def \Kbar{\bar K}
\def\bd{B_d^0}
\def\bs{B_s^0}
\def\btod{{\bar b} \to {\bar d}}
\def\btos{{\bar b} \to {\bar s}}
\def\ANPq{{\cal A}^q}
\def\ApNPqph{{\cal A}^{\prime,q} e^{i \Phi'_q}}
\def\ApNPCqph{{\cal A}^{\prime {\sss C}, q} e^{i \Phi_q^{\prime C}}}
\def\ApNPCuph{{\cal A}^{\prime {\sss C}, u} e^{i \Phi_u^{\prime C}}}
\def\ApNPCdph{{\cal A}^{\prime {\sss C}, d} e^{i \Phi_d^{\prime C}}}
\def\pewcp{P_{\sss EW}^{\prime\sss C}}
\def\pewp{P'_{\sss EW}}
\def\ApNPuph{{\cal A}^{\prime,u} e^{i \Phi'_u}}
\def\ApNPdph{{\cal A}^{\prime,d} e^{i \Phi'_d}}
\def\ApNPcomb{{\cal A}^{\prime, comb} e^{i \Phi'}}
\def\btopik{B \to \pi K}

\begin{document}

\begin{flushright}
UdeM-GPP-TH-07-157
\end{flushright}

\begin{center}
\bigskip
{\Large \bf U-Spin Tests of the Standard Model and New
Physics}\\
\bigskip
\bigskip
{\large
Makiko Nagashima\footnote{makiko@lps.umontreal.ca},
Alejandro Szynkman\footnote{szynkman@lps.umontreal.ca}
and David London\footnote{london@lps.umontreal.ca}, \\
}
\end{center}

\begin{flushleft}
~~~~~~~~~~~~~~~{\it Physique des Particules, Universit\'e
de Montr\'eal,}\\
~~~~~~~~~~~~~~~{\it C.P. 6128, succ. centre-ville, Montr\'eal, QC,
Canada H3C 3J7}\\
\end{flushleft}

\begin{center}
\bigskip (\today)
\vskip0.5cm {\Large Abstract\\} \vskip3truemm
\parbox[t]{\textwidth}{Within the standard model, a relation
involving branching ratios and direct CP asymmetries holds
for the $B$-decay pairs that are related by U-spin. The
violation of this relation indicates new physics (NP). In
this paper, we assume that the NP affects only the $\Delta S
= 1$ decays, and show that the NP operators are generally the
same as those appearing in $\btopik$ decays.  The fit to the
latest $\btopik$ data shows that only one NP operator is
sizeable. As a consequence, the relation is expected to be
violated for only one decay pair: $\bd \to K^0\pi^0$ and $\bs
\to \Kbar^0\pi^0$.}
\end{center}

\thispagestyle{empty}
\newpage
\setcounter{page}{1}
\baselineskip=14pt

At present, several different experiments are focusing on the
measurement of CP violation in various $B$ decays. The
principal hope is to find a discrepancy with the predictions
of the standard model (SM). This would indicate the presence
of physics beyond the SM, which is one of the main aims of
these experiments.

U-spin relations between different $B$ decays provide several
tests of the SM. This is discussed in detail in
Refs.~\cite{Gronau,Gronau2}. U-spin is the symmetry that
places $d$ and $s$ quarks on an equal footing, and is often
given as transposing $d$ and $s$ quarks: $d \leftrightarrow
s$. Assuming a perfect U-spin symmetry, the effective
Hamiltonian describing a $\Delta S = 0$ transition ($\btod$)
is equal to that of the corresponding $\Delta S = 1$
transition ($\btos$) with $d \leftrightarrow s$ (the elements
of the Cabibbo-Kobayashi-Maskawa (CKM) matrix are changed
appropriately). Using the CKM unitarity relation \cite{Jarl},
\beq
{\rm Im}(V^*_{ub}V_{us}V_{cb}V^*_{cs}) = - {\rm
Im}(V^*_{ub}V_{ud}V_{cb}V^*_{cd}) ~,
\eeq
this implies that there exists a U-spin relation between the
CP-violating rate differences of the $\Delta S = 0$ and
$\Delta S = 1$ decays \cite{Gronau,Gronau2}:
\beq
\vert A(B \to f)\vert^2 - \vert A(\Bbar \to \fbar)\vert^2 \\
= -\left[ \vert A(UB \to Uf)\vert ^2 - \vert A(U\Bbar \to U\fbar) \vert^2 \right] ~, \nn
\eeq
in which $U$ is the U-spin operator that transposes $d$ and
$s$ quarks. This expression can be written as
\beq
{- A_{\sss CP}^{dir}{\hbox{(decay \#1)}} \over A_{\sss CP}^{dir}{\hbox{(decay \#2)}}} =
{BR{\hbox{(decay \#2)}} \over BR{\hbox{(decay \#1)}}} ~,
\label{Uspinreln}
\eeq
where $A_{\sss CP}^{dir}$ and $BR$ refer to the direct CP
asymmetry and branching ratio, respectively, and where decays
\#1,2 are the $\Delta S = 0$ and $\Delta S = 1$ decays, in
either order, related by U-spin. Note that if decays \#1,2
include $\bd$ and $\bs$ mesons, there is an additional factor
on the right-hand side taking the lifetime difference into
account.

The pairs of $B\to PP$ decays ($P$ is a pseudoscalar meson)
which are related by U-spin are
\begin{enumerate}
\item $\bd \to K^+\pi^-$ and $\bs \to \pi^+ K^-$~,
\item $\bs \to K^+ K^-$ and $\bd \to \pi^+\pi^-$~,
\item $\bd \to K^0\pi^0$ and $\bs \to \Kbar^0\pi^0$~,
\item $B^+ \to K^0 \pi^+$ and $B^+ \to \Kbar^0 K^+$~,
\item $\bs \to K^0 \Kbar^0$ and $\bd \to \Kbar^0 K^0$~,
\item $\bs \to\pi^+\pi^-$ and $\bd \to K^+ K^-$~.
\end{enumerate}
In all cases, the first decay is $\Delta S = 1$; the second is $\Delta
S = 0$. 

Note that the decays described in pair \#6 can only come
about if the two quarks in the initial state interact with
each other (annihilation). We therefore expect the branching
ratios for these decays to be considerably smaller than those
of the other decays. For this reason, we ignore this pair
from here on.

Throughout this paper, we refer to a given pair of decays by
its number in the list above (\#1--\#5). If
Eq.~\ref{Uspinreln} is not satisfied for any of these pairs
of decays, this indicates the presence of physics beyond the
SM.

Technically, this is not quite true, as U-spin is only an
approximate symmetry. It is a quite common assumption
(motivated by factorization in tree-level amplitudes)
\cite{Chiang} to take U-spin breaking effects in branching
ratios to be given by (known) ratios of decay constants. In
this case, the effects total $O(30\%)$. If one includes form
factors, QCD sum rules give U-spin breaking effects of
$O(80\%)$ \cite{QCDsumrules}. However, both of these
estimates are somewhat misleading -- they imply that the {\it
uncertainty} on the U-spin breaking is large, assuming that
the decays related by U-spin are unknown. For example, one
doesn't know if the decay-constant breaking effect is
$f_{\sss K}/f_\pi$ or $f_\pi/f_{\sss K}$. However, large
U-spin breaking is not a problem for a given decay pair as
long as (i) the U-spin breaking is taken into account, and
(ii) the uncertainty is not too big.

In this paper, we {\it know} which decays are involved. We
can thus compute the U-spin breaking effect within naive
factorization. The main descriptions of hadronic $B$ decays
(soft collinear effective theory )SCET) \cite{SCET}, QCD
factorization (QCDf) \cite{BBNS,BBNS2,BBNS3}, perturbative
QCD (pQCD) \cite{pQCD}) all include an expansion in powers of
$1/m_b$. Within QCDf it is proven that factorization
corresponds to the leading-order term. Thus naive
factorization is a good approximation, and nonfactorizable
effects less important, if the higher-order terms in $1/m_b$
are small. One example of a large such term, which would
invalidate this approximation, is penguin annihilation
\cite{Kagan}, proposed to explain the $B\to\phi K^*$
polarization puzzle. However, to date there is no firm
evidence that penguin annihilation is large. Other examples
of higher-order terms are exchange and annihilation diagrams,
and there is no experimental evidence that these are
sizeable. Thus, at present the evidence points to small
higher-order terms, and naive factorization is a reasonable
approximation. For U-spin breaking we find
\begin{enumerate}
\item $\Delta U = (f_{\sss K} F_{\bd \to \pi}) / (f_\pi F_{\bs \to
K}) = 1.02 \pm 0.11$,
\item $\Delta U = (f_{\sss K} F_{\bs \to K}) / (f_\pi F_{\bd \to
\pi}) = 1.49 \pm 0.15$,
\item $\Delta U = (f_{\sss K} F_{\bd \to \pi}) / (f_\pi F_{\bs \to
K}) = 1.02 \pm 0.11$,
\item $\Delta U = F_{B^+ \to \pi} / F_{B^+ \to K} = 0.79 \pm 0.04$, 
\item $\Delta U = F_{\bs \to K} / F_{\bd \to K} = 0.95 \pm 0.11$,
\end{enumerate}
where $f_i$ is the decay constant and $F_i$ is the form
factor.  In working out the numerical values for the $\Delta
U$'s, we take $f_\pi = 130$ MeV, $f_{\sss K} = 160$ MeV,
$F_{\bs \to K} / F_{\bd \to \pi} = {1.21}^{+0.14}_{-0.11}$
\cite{khodj1}, $F_{B^+\to K}/F_{B^+\to \pi} = 1.27 \pm 0.07$
\cite{khodj2}. For the form-factor ratio in \#5, we take
$F_{\bd\to \pi}/F_{\bd\to K} = F_{B^+ \to \pi} / F_{B^+ \to
K}$ and combine this with $F_{\bs \to K} / F_{\bd \to
\pi}$. The calculation of $\Delta U$ for \#3 is complicated
by the fact that there are several contributing diagrams,
with different naive U-spin breaking effects. In order to
obtain a numerical result, we assume that $f_{\sss K}/f_\pi
\simeq F_{\bd\to K}/F_{\bd\to \pi}$, which is confirmed by
Refs.~\cite{khodj1,khodj2}. Note that the central values give
small U-spin breaking for pairs \#1, \#3 and \#5. The
uncertainty on the $\Delta U$'s is then due to extent to
which the form factors are known. In addition, there are
nonfactorizable effects [$O(1/m_b)$], which we estimate to be
$\sim 10\%$. The total error on U-spin breaking is then
relatively small, $O(20\%)$.

It is worth mentioning that, in the case of pairs \#1 and
\#3, we expect U-spin breaking effects beyond naive
factorization to have a small impact since the final states
for both $\bd$ and $\bs$ decays in each pair are charge
conjugate \cite{Lipkin}.

Eq.~\ref{Uspinreln} must now be modified to include the above
U-spin breaking effects. We find that the modification is
simple:
\beq
{- A_{\sss CP}^{dir}{\hbox{(decay \#1)}} \over A_{\sss CP}^{dir}{\hbox{(decay \#2)}}} =
\Delta U^2 {BR{\hbox{(decay \#2)}} \over BR{\hbox{(decay \#1)}}} ~,
\label{Uspinreln2}
\eeq
If the above equation is not satisfied (including the error
on $\Delta U$) for any of the U-spin pairs of decays, this
indicates the presence of physics beyond the SM.

Taking the U-spin breaking into account, and given that the
error is not very large, one will not have a gross violation
of Eq.~\ref{Uspinreln2}.  Thus, it is more correct to say
that ``if Eq.~\ref{Uspinreln2} is greatly violated for any of
the pairs of decays -- such as the direct CP asymmetries
having the same sign -- this indicates the presence of
physics beyond the SM.''

This new physics (NP) can take the form of new contributions
to the $\Delta S = 0$ and/or the $\Delta S = 1$ decays. Now,
there have been many measurements of quantities in $B$
decays. To date, there have been several hints of
discrepancies with the SM. However, none of them lie in the
$\Delta S = 0$ sector; they all point to NP in $\btos$
transitions. For example, the CP asymmetry in $\btos q{\bar
q}$ modes ($q=u,d,s$) is found to differ from that in ${\bar
b} \to {\bar c} c {\bar s}$ decays by 2.6$\sigma$ (they are
expected to be approximately equal in the SM)
\cite{HFAG,Zhou,Zhou2}. One also sees a discrepancy with the
SM in triple-product asymmetries in $B \to \phi K^*$
\cite{BVVTP,phiKstarTP,phiKstarTP2}, and in the polarization
measurements of $B \to \phi K^*$
\cite{BphiKstar_exp,BphiKstar_exp2,BphiKstarNP,BphiKstarNP2,BphiKstarNP3,
BphiKstarNP4,BphiKstarNP5,BphiKstarNP6,BphiKstarNP7,
BphiKstarSM, BphiKstarSM2, BphiKstarSM3, BphiKstarSM4,
BphiKstarSM5} and $B\to \rho K^*$
\cite{BrhoKstar_exp,BrhoKstar_exp2,BrhoKstar_exp3,BrhoKstar}.
Finally, some $B\to\pi K$ measurements disagree with SM
expectations \cite{Baekandquim,Baekandquim2,Baekandquim3},
although it has been argued that the so-called $\btopik$
puzzle \cite{BFRS,BFRS2,BFRS3} has been somewhat reduced
\cite{puzzle,puzzlea,puzzleb,puzzle2,puzzle2a}.  Although
none of the discrepancies are statistically significant,
together they give an interesting hint of NP.  In this paper
we follow this indication and assume that the NP appears only
in $\btos$ decays ($\Delta S = 1$) but does not affect
$\btod$ decays ($\Delta S = 0$).

There are many NP operators which can contribute to $\Delta S
= 1$ decays. However, it was recently shown in Ref.~\cite{DL}
that this number can be reduced considerably. Briefly, the
argument is as follows. Following the experimental hints, we
assume that NP contributes significantly to those decays
which have large $\btos$ penguin amplitudes, and take the NP
operators to be roughly the same size as the SM $\btos$
penguin operators, so the new effects are sizeable. Each NP
matrix element can have its own weak and strong phase.  Now,
all strong phases arise from rescattering. In the SM, this
comes mainly from the $\btos c {\bar c}$ tree diagram. The NP
strong phases must come from rescattering of the NP
operators.  However, the tree diagram is quite a bit larger
than the $\btos$ penguin diagram (the expected size of the NP
operator). As a consequence, the generated NP strong phases
are correspondingly smaller than those of the SM. That is,
the NP strong phases are negligible compared to the SM strong
phases. (The appendix of Ref.~\cite{DLmethods} contains a
detailed discussion of the small NP strong phases.)

The idea of small NP strong phases is central to the analysis
presented in this paper. As such, it is worthwhile to examine
the circumstances under which the idea holds or fails. NP
strong phases can be mimicked by rescattering from operators
that are higher order in $1/m_b$. Thus, the idea of small NP
strong phases can fail if higher-order terms in $1/m_b$ are
large, i.e.\ if the coefficients of such terms are bigger
than expected. However, as discussed above in the context of
U-spin breaking, there is no evidence that these terms are
sizeable. Thus, although the idea of small NP strong phases
can fail if higher-order terms in $1/m_b$ are large, at
present these terms appear to be small. We therefore conclude
that the idea of small NP strong phases is justified.

The neglect of all NP strong phases considerably simplifies
the situation. At the quark level, each NP contribution to
the decay $B\to f$ takes the form $\bra{f} {\cal O}_{\sss
NP}^{ij,q} \ket{B} $, where ${\cal O}_{\sss NP}^{ij,q} \sim
{\bar s} \Gamma_i b \, {\bar q} \Gamma_j q$ ($q = u,d,s,c$),
in which the $\Gamma_{i,j}$ represent Lorentz structures, and
colour indices are suppressed. If one neglects all NP strong
phases, one can now combine all NP matrix elements into a
single NP amplitude, with a single weak phase:
\beq
\sum \bra{f} {\cal O}_{\sss NP}^{ij,q} \ket{B} = \ANPq e^{i \Phi_q} ~.
\eeq

Thus, all NP effects can be parametrized in terms of a small
number of NP quantities. For $\Delta S =1$ decays, there are
two classes of NP operators, differing in their colour
structure: ${\bar s}_\alpha \Gamma_i b_\alpha \, {\bar
q}_\beta \Gamma_j q_\beta$ and ${\bar s}_\alpha \Gamma_i
b_\beta \, {\bar q}_\beta \Gamma_j q_\alpha$. The first class
of NP operators contributes with no colour suppression to
final states containing ${\bar q}q$ mesons. (The second type
of operator can also contribute via Fierz transformations,
but there is a suppression factor of $1/N_c$, as well as
additional operators involving colour octet currents.)
Similarly, for final states with ${\bar s} q$ mesons, the
roles of the two classes of operators are reversed. As in
Ref.~\cite{DLmethods}, we denote by $\ApNPqph$ and
$\ApNPCqph$ the sum of NP operators which contribute to final
states involving ${\bar q}q$ and ${\bar s}q$ mesons,
respectively (the primes indicate a ($\Delta S =1$) $\btos$
transition). Here, $\Phi'_q$ and $\Phi_q^{\prime {\sss C}}$
are the NP weak phases; the strong phases are zero. We stress
that, despite the ``colour-suppressed'' index $C$, the
operators $\ApNPCqph$ are not necessarily smaller than the
$\ApNPqph$.

It is now possible to easily compute the effect of the NP
operators on the $\Delta S =1$ decays mentioned in this
paper. However, before doing so, we return to $B\to \pi K$
decays. In Ref.~\cite{BL} a fit was done to the 2006
$\btopik$ data, shown in Table 1. We summarize the results
here. If one defines the ratios
\begin{eqnarray}
{R}=\frac{\tau_{B^+}}{\tau_{B^0}}
    \frac{BR[B^0\to\pi^-K^+]+BR[\bar B^0\to\pi^+K^-]}
        {BR[B_d^+\to\pi^+K^0]+BR[B_d^-\to\pi^-\bar K^0]} ~, \nonumber\\
 {R_n}=\frac12 \frac{BR[B^0\to\pi^-K^+]+BR[\bar B^0\to\pi^+K^-]}
        {BR[B^0\to\pi^0K^0]+BR[\bar B^0\to\pi^0 \bar K^0]} ~, \nonumber\\
 {R_c}=2\frac{BR[B_d^+\to\pi^0 K^+]+BR[B_d^-\to\pi^0 K^-]}
        {BR[B_d^+\to\pi^+ K^0]+BR[B_d^-\to\pi^- \bar K^0]} ~,
\end{eqnarray}
one can show that the present values of $R$, $R_n$ and $R_c$
agree with the SM \cite{hurth}. This has led some authors to
posit that there is no longer a $B \to \pi K$ puzzle
\cite{hurth,nopuzzle}.  Unfortunately, this analysis is
incomplete.

\begin{table}[tbh]
\center
\begin{tabular}{cccc}
\hline
\hline
Mode & $BR[10^{-6}]$ & $A_{\sss CP}$ & $S_{\sss CP}$ \\ \hline
$B^+ \to \pi^+ K^0$ & $23.1 \pm 1.0$ & $0.009 \pm 0.025$ & \\
$B^+ \to \pi^0 K^+$ & $12.8 \pm 0.6$ & $0.047 \pm 0.026$ & \\
$\bd \to \pi^- K^+$ & $19.7 \pm 0.6$ & $-0.093 \pm 0.015$ & \\
$\bd \to \pi^0 K^0$ & $10.0 \pm 0.6$ & $-0.12 \pm 0.11$ &
$0.33 \pm 0.21$ \\
\hline
\hline
\end{tabular}
\caption{Branching ratios, direct CP asymmetries $A_{\sss
CP}$, and mixing-induced CP asymmetry $S_{\sss CP}$ (if
applicable) for the four $\btopik$ decay modes. The data is
taken from
Refs.~\cite{HFAG,piKrefs,piKrefs2,piKrefs3,piKrefs4,piKrefs5,piKrefs6,piKrefs7,piKrefs8,
piKrefs9}.}
\label{tab:data}
\end{table}

In Refs.~\cite{GHLR,GHLR2}, the relative sizes of the SM
$\btopik$ diagrams were roughly estimated as
\beq
1 : |P'_{tc}| ~~,~~~~ {\cal O}({\bar\lambda}) : |T'|,~|\pewp| ~~,~~~~
{\cal O}({\bar\lambda}^2) : |C'|,~|P'_{uc}|,~|\pewcp| ~, 
\label{ampsizes}
\eeq
where ${\bar\lambda} \sim 0.2$. These estimates are expected
to hold approximately in the SM. Now, one can show that the
amplitudes $\sqrt{2} A(B^+ \to \pi^0 K^+)$ and $A(\bd \to
\pi^- K^+)$ are equal, up to a factors of the small diagrams
$\pewp$ and $C'$. We therefore expect $A_{\sss CP}(B^+ \to
\pi^0 K^+)$ to be approximately equal to $A_{\sss CP}(\bd \to
\pi^- K^+)$ in magnitude (multiplicative factors cancel
between the numerator and denominator of
asymmetries). However, as can be seen in Table 1, these
asymmetries are very different. Thus, the present $\btopik$
data cannot be explained by the SM. It is only by considering
the CP-violating asymmetries that one realizes this.

If one includes all diagrams, a good fit is found
\cite{BL}. This has led some people to argue that there is no
discrepancy in $\btopik$ decays \cite{CKMfitter}\footnote{
According to its website, it appears that the CKMfitter Group
has modified its point of view compared with this paper. The
$\btopik$ measurements are now not part of the global fit to
determine the CKM parameters. In particular, the various
$\sin 2\beta^{\rm eff}$ results from penguin-dominated decays
are not included. This suggests that the CKMfitter Group
feels that there are some curious results in $\btos$
transitions.}  . However, this fit requires $|C'/T'| = 1.6
\pm 0.3$. This value is much larger than the naive estimates
of Eq.~(\ref{ampsizes}), the NLO pQCD prediction, $|C'/T'|
\sim 0.3$ \cite{PQCD_NLO}, and the maximal SCET (QCDf)
prediction, $|C'/T'| \sim 0.6$
\cite{SCET,Beneke:2003zv}. Thus, if one takes these
theoretical results seriously, one is led to conclude that
the $\btopik$ puzzle is still present, at the level of $\gsim
3\sigma$.  Indeed, the puzzle is much worse in 2006 than in
2004 \cite{Baeketal}.

Assuming that the effect is not a statistical fluctuation,
one must add NP operators. If one ignores the small ${\cal
O}({\bar\lambda}^2)$ diagrams, the $B\to \pi^i K^j$
amplitudes ($i,j$ are electric charges) can be written
\cite{DLmethods}
\bea
\label{piKamps}
A^{+0} &\!\!=\!\!& -P'_{tc} + \ApNPCdph ~, \nn\\
\sqrt{2} A^{0+} &\!\!=\!\!& P'_{tc} - T' \, e^{i\gamma} -
\pewp + \ApNPcomb - \ApNPCuph ~, \nn\\
A^{-+} &\!\!=\!\!& P'_{tc} - T' \, e^{i\gamma} - \ApNPCuph
~, \nn\\
\sqrt{2} A^{00} &\!\!=\!\!& -P'_{tc} - \pewp
+ \ApNPcomb + \ApNPCdph ~,
\eea
where $\ApNPcomb \equiv - \ApNPuph + \ApNPdph$. It is not
possible to distinguish the two component amplitudes in
$\btopik$ decays.  $\gamma$ is the SM weak phase.

The NP operators mentioned above correspond to the decay $B
\to \pi K$.  But the other $\Delta S = 1$ decays in the
U-spin list include $\bs \to K^+ K^-$ and $\bs \to K^0
\Kbar^0$. One might think that the NP operators affecting
these decays bear no relation to those in $\btopik$. In fact,
this is not true: the other $\Delta S = 1$ decays are the
same as $\btopik$ in the limit of flavour SU(3) (which treats
$u$, $d$ and $s$ quarks identically).  The point is that the
matrix elements differ only in the quarks involved, which
affects their hadronization. If all quarks are identical
[flavour SU(3)], then the hadronization is the same (the NP
affects this hadronization only at the level of $m_b/M_{\sss
NP}$, which is tiny). Thus, the NP operators for all $\Delta
S = 1$ decays are the same, up to SU(3)-breaking effects. We
will therefore denote all NP operators as $\ApNPCuph$,
$\ApNPCdph$, and $\ApNPcomb$.

This is the first main result of this paper. Within SU(3),
all $\Delta S = 1$ decays receive contributions from the same
NP operators.  Constraints on most of these operators can be
taken from the analysis of $\btopik$ decays.

We can now compute the contribution of NP operators to all
$\Delta S = 1$ decays in the U-spin list. $\bd \to K^+\pi^-$
(pair \#1) and $\bs \to K^+ K^-$ (pair \#2) receive a NP
contribution of the form $\ApNPCuph$; $\bd \to K^0\pi^0$
(pair \#3) receives $\ApNPcomb + \ApNPCdph$; $B^+ \to K^0
\pi^+$ (pair \#4) and $\bs \to K^0 \Kbar^0$ (pair \#5)
receive $\ApNPCdph$. Depending on the expected size of the NP
operators, not all decays will be equally affected by the
NP. In particular, if a given NP operator is expected to be
small, Eq.~\ref{Uspinreln2} will be satisfied for this pair
of decays. Conversely, if Eq.~\ref{Uspinreln2} is violated
for a particular decay pair, this points to the presence of a
specific NP operator.

The three NP operators were then included in the $\btopik$
fit in Ref.~\cite{BL}, one at a time. It was found that the
fit remained poor if $\ApNPCuph$ or $\ApNPCdph$ was
added. That is, large values of these NP operators may be
allowed, but there is no experimental evidence that these are
needed. Below we therefore assume that $\ApNPCuph$ and
$\ApNPCdph$ are small. It was also found that the discrepancy
could be removed if the NP operator $\ApNPcomb$ were added.
That is, a large value of $\ApNPcomb$ was implied. In
summary, the present $\btopik$ puzzle suggests that the NP
operators $\ApNPCuph$ and $\ApNPCdph$ are small, but that
$\ApNPcomb$ is big. This is what was found previously
\cite{Baeketal}.

In this case, only the decay $\bd \to K^0\pi^0$ (pair \#3)
can be significantly affected by the NP. We therefore
conclude that the present $\btopik$ data predicts that, of
the five U-spin pairs, one expects a measurable discrepancy
with the SM (taking U-spin breaking into account) only for
pair \#3: $\bd \to K^0\pi^0$ and $\bs \to \Kbar^0\pi^0$. This
is the second main result of this paper.

Note that this result is based on the conclusion that a
$\btopik$ puzzle is present. However, suppose that all
discrepancies in the $\btopik$ system disappear. In this
case, all NP operators are small, which means that there will
be no disagreement with the SM for any U-spin pair.

If $\pi K$ final-state rescattering is sizeable, the
transitions $K^+\pi^- \leftrightarrow K^0\pi^0$ and $\pi^+
K^- \leftrightarrow \Kbar^0\pi^0$ can be large, which implies
that pair \#1 can be generated by rescattering from pair \#3.
In this case, U-spin breaking in pair \#1 can come from
U-spin breaking in pair \#3, and can be big even if
$\ApNPCuph$ is small.  On the other hand, $\pi K$ final-state
rescattering is higher order in the $1/m_b$ expansion of QCDf
and pQCD, and is thus expected to be small. If this
theoretical input is valid, such rescattering is negligible,
and the conclusion holds that there is a discrepancy with the
SM only for pair \#3.  However, if it turns out that $\pi K$
final-state rescattering effects are in fact large, then we
could expect significant U-spin breaking in pair \#1 as well
(but see below).

In general, measurements have not yet been made to test the
U-spin relation between pairs of $B$ decay. The one exception
is pair \#1.  Recently, the branching ratio and direct CP
asymmetry of $\bs \to \pi^+ K^-$ have been measured by the
CDF experiment \cite{CDFrecent}: $BR(\bs \to\pi^+
K^-)=(5.00\pm 0.75\pm 1.00)\times 10^{-6}$ and $A_{\sss
CP}^{dir}=0.39\pm 0.15\pm 0.08$.  Together with the
experimental measurements of $\bd \to \pi^- K^+$ shown in
Table~1, we have
\beq
{- A_{\sss CP}^{dir}(\bs \to \pi^+ K^-) \over 
   A_{\sss CP}^{dir}(\bd \to K^+ \pi^-)} = 4.2 \pm 2.0 ~~,~~~~
{BR(\bd \to K^+\pi^-) \over BR(\bs \to\pi^+ K^-)} = 3.9 \pm 1.0 ~.  
\eeq
We therefore see that, although the errors are still large,
the two ratios are equal, and that Eq.~\ref{Uspinreln2} is
satisfied by the current data.  This suggests that $\pi K$
final-state rescattering is in fact small (or that there are
cancellations with the NP).

In addition, the SM predicts that $BR(\bs \to\pi^+ K^-)\sim
(3$--$10) \times 10^{-6}$
\cite{SMprdBspiK,SMprdBspiK2,SMprdBspiK3}, in agreement with
measurement.  Moreover, recent theoretical calculations
within the SM \cite{piKSM} make predictions for $\bd \to
\pi^- K^+$ which are consistent with the results shown in
Table 1. This indicates that pair \#1 shows no sign of NP.
In other words, it supports the idea that $\ApNPCuph$ is
small.

Finally, above we showed that present $\btopik$ data predicts
a disagreement with Eq.~\ref{Uspinreln2} only for pair \#3
($\bd \to K^0\pi^0$ and $\bs \to \Kbar^0\pi^0$). Obviously,
it will be important to check U-spin violation in all the
$B$-decay pairs, but special attention will be paid to this
pair, since a nonzero effect is expected. In particular, it
will be necessary to measure the branching ratio and direct
CP asymmetry in $\bs \to \Kbar^0\pi^0$. This will be
challenging.

In fact, measurements of $\bs \to \Kbar^0\pi^0$ are quite
impracticable at present colliders. Since the $\bs$ direction
cannot be determined at hadron colliders, the $\bs$ decay
vertex cannot be reconstructed via $\ks\to \pi^+\pi^-$
($\pi^0\to \gamma\gamma$ leaves no track). Instead, with a
similar technique to that used for $\bd\to K^0 \pi^0$
measurements, the ({\it Super}) $B$ factory must be used for
measurements of this decay.  However, the direct CP asymmetry
in $\Kbar^0\pi^0$ requires measurements of the time-dependent
CP asymmetry. For this purpose, {\it Super} $B$ must have a
much better $\Delta t$ resolution than planned (an
improvement of an order of magnitude), so that $\Delta
m_{\bs}$ measurements can be performed.  It is ironic that,
although most $\bs$ decays can be well studied by hadron
collider experiments, the best machine for a U-spin test
turns out to be the {\it Super} $B$ factory with a better
$\Delta t$ resolution.

In summary, some time ago it was pointed out that, within the
standard model (SM), a relation involving branching ratios
and direct CP asymmetries [Eq.~\ref{Uspinreln2}] holds for
two $B$ decays that are related by U-spin. This equation
includes U-spin breaking.  There are five (non-annihilation)
decay pairs to which this applies. If this relation is found
not to hold for a given pair, this implies the presence of
physics beyond the SM in that pair.  In this paper, we follow
the experimental indications and assume that this new physics
(NP) appears only in $\btos$ decays ($\Delta S = 1$) but does
not affect $\btod$ decays ($\Delta S = 0$). There are only a
handful of NP operators that can affect the $\Delta S = 1$
$B$-decay amplitudes. We have shown that, to a good
approximation, these operators are the same as the three
appearing in $\btopik$ decays. The fit to the latest
$\btopik$ data shows that only one NP operator is found to be
large. As a result, Eq.~\ref{Uspinreln2} is expected to be
violated by only one decay pair: $\bd \to K^0\pi^0$ and $\bs
\to \Kbar^0\pi^0$. The measurement of the violation of
Eq.~\ref{Uspinreln2} in this $B$-decay pair will thus be a
test of NP in $\btopik$ decays.

\bigskip
\noindent
{\bf Acknowledgments}:
We thank Seungwon Baek, Alakabha Datta and Fumihiko Ukegawa
for very helpful communications. This work was financially
supported by NSERC of Canada.


\end{document}